\documentstyle[12pt,a4]{article}
\catcode `@=11 \@addtoreset{equation}{section} \catcode `@=12
\input epsf
\begin{document}
\def\be{\begin{equation}}
\def\ee{\end{equation}}
\def\beq{\begin{eqnarray}}
\def\eeq{\end{eqnarray}}
\def\und{\underbar}
\def\T{{\cal T}}
\def\eq#1{(\ref{#1})}
\def\l{\label}
\begin{flushright}
CERN-TH/95-186 \\
TIFR/TH/95-30
\end{flushright}
\vspace{1 ex}
\begin{center}
{\Large{\bf DISCRETE-STATE MODULI OF  }} \\
\vspace{1 ex}
{\Large{\bf STRING THEORY FROM }} \\
\vspace{1 ex}
{\Large{\bf THE C=1 MATRIX MODEL}} \\
\vspace{7 ex}
{\large Avinash Dhar}\footnote{On leave from Theoretical Physics
Group, Tata Institute of Fundamental Research, Homi Bhabha Road,
Bombay 400 005, INDIA.},\footnote{e-mail: adhar@surya11.cern.ch} \\
\vspace{2 ex}
Theory Division, CERN, CH-1211, Geneva 23, SWITZERLAND \\
\vspace{3 ex}
{\large Gautam Mandal}\footnote{e-mail: mandal@theory.tifr.res.in} and
{\large Spenta R. Wadia}\footnote{e-mail: wadia@theory.tifr.res.in} \\
\vspace{2 ex}
Tata Institute of Fundamental Research \\
Homi Bhabha Road, Bombay 400 005, INDIA \\
\vspace{5 ex}
{\large{\underbar{ABSTRACT}}} \\
\end{center}
\vspace{3 ex}
We propose a new formulation of the space-time interpretation of the
$c=1$ matrix model.  Our formulation uses the well-known leg-pole
factor that relates the matrix model amplitudes to that of the
2-dimensional string theory, but includes fluctuations around the
fermi vacuum on {\sl both sides} of the inverted harmonic
oscillator potential of the double-scaled model, even when the
fluctuations are small and confined entirely within the asymptotes in
the phase plane.  We argue that including fluctuations on both sides
of the potential is essential for a consistent interpretation of the
leg-pole transformed theory as a theory of space-time gravity. We
reproduce the known results for the string theory tree level
scattering amplitudes for the tachyon in flat space and linear dilaton
background as
a special case.  We show that the generic case corresponds to more
general space-time backgrounds.  In particular, we identify the
parameter corresponding to the background metric perturbation in string
theory (black hole mass) in terms of the matrix model variables.
Possible implications of our work for a consistent nonperturbative
definition of string theory as well as for quantized gravity and
black-hole physics are discussed.
\vfill\eject
\renewcommand{\Large}{\large}
\renewcommand{\LARGE}{\large}

\section{Introduction and Summary}

It is well-known that the scattering amplitudes of the scalar
excitation of the double-scaled $c=1$ matrix model are not identical
to the tachyon scattering amplitudes of 2-dimensional string theory,
but are related to these by a `leg-pole' factor \cite{one}.  Although
this leg-pole factor is a pure phase in momentum space, it translates
into a nonlocal renormalization of the wavefunctions of the scalar
excitation of the matrix model, and gives rise to all of space-time
gravitational physics of the string theory \cite{two}, which is
otherwise absent in the matrix model.

The $c=1$ matrix model is equivalent to a theory of nonrelativistic,
noninteracting fermions in an inverted harmonic oscillator potential
in the double scaling limit \cite{three}.  The semiclassical physics
of the matrix model is, therefore, described by a fermi liquid theory.
The existing space-time interpretation of the matrix model is based on
small fluctuations of this fermi fluid, around the fermi vacuum, on
only one side of the potential (Figure 1).

\begin{figure}

\epsfbox{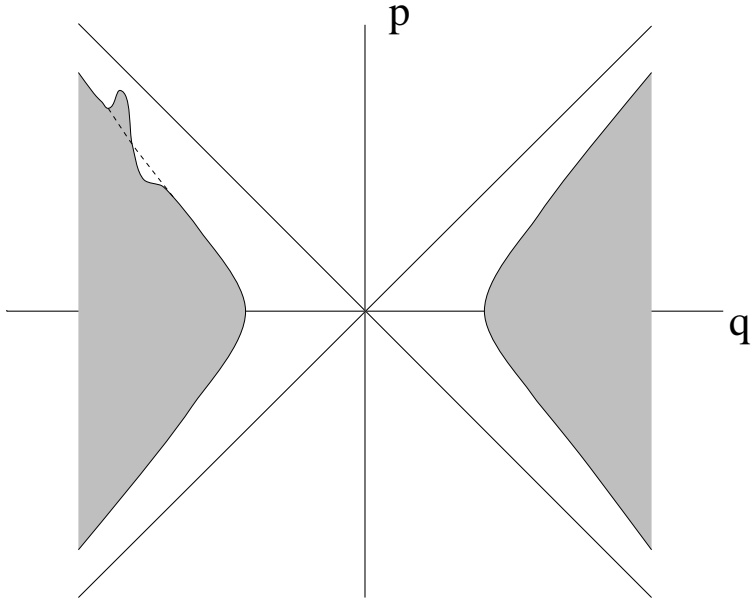}

\caption{Small fluctuation of the fermi surface on only one
side of the potential.}

\end{figure}

It is generally believed that the other side can be ignored as long as
one avoids the classical configurations consisting of large
fluctuations in which a part of the fluid croses the asymptotes (Figure
2),
\begin{figure}

\epsfbox{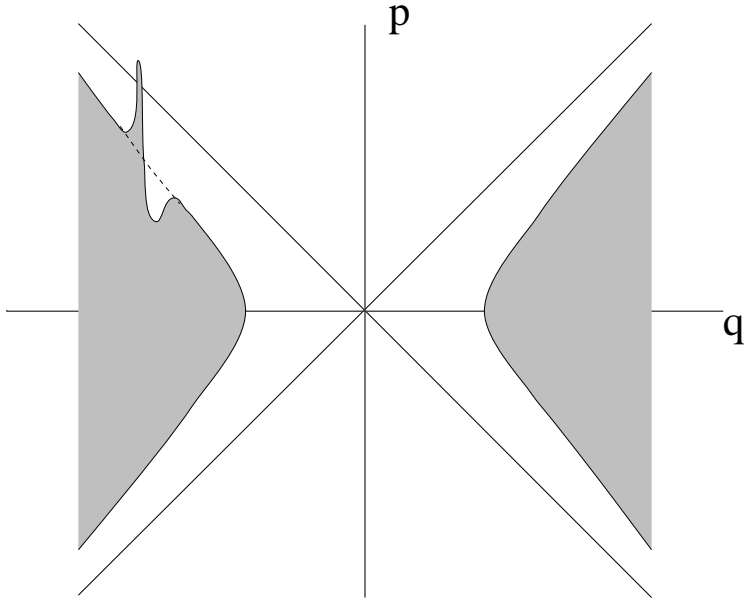}

\caption{A part of the fluid fluctuation crosses the $p=-q$ asymptote.}

\end{figure}
and nonperturbative tunnelling issues \cite{four}.

There is, however, an argument which suggests that one has to decide
the fate of the other side of the potential even in the small-field
perturbative situation when the two sides of the potential are
entirely decoupled in the matrix model.  The point is that string
theory is a theory of gravity, and in any consistent space-time
interpretation of the matrix model the space-time metric must couple
to the
\underbar{entire} energy-momentum tensor of the theory.  Now, a
generic perturbation of the fermi fluid has fluctuations on
\underbar{both} sides of the potential, even in the small-field
perturbative situation, and the total energy of this configuration has
contributions from fluctuations on both sides.  \underbar{The string
theory metric must couple to this total} \underbar{energy}, since the
hamiltonian of the string theory is identical to that of the matrix
model, unless we decide to remove the other side from the start, by
modifying the potential.  This is presumably also the case for all the
other higher conserved charges corresponding to the diagonal
generators of $W$-infinity symmetry of the matrix model.  Thus, we
must decide on the fate of the other side of the potential, for
\underbar{any} fermi fluid configuration, before arriving at a
consistent space-time interpretation of the model.

In this paper we will retain both sides of the potential and consider
the generic situation in which there are fluctuations of the fermi
surface on both sides (Figure 3).
\begin{figure}

\epsfbox{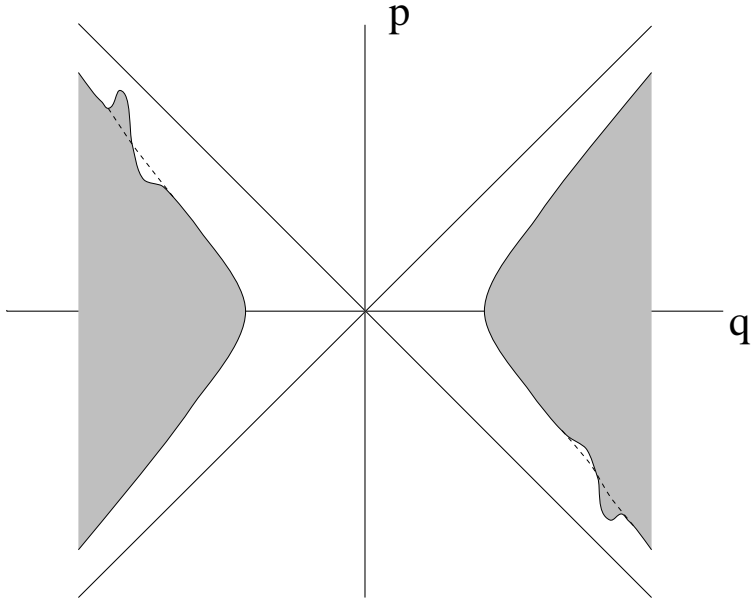}

\caption{A generic initial small-field fluctuation of the fermi
surface.}

\end{figure}
The summary of our results is as follows. The usual leg-pole transform
of the \underbar{special configurations} in which the fluctuations of
the fermi surface, lying entirely within the asymptotes, are identical
on the two sides, reproduces the tachyon scattering amplitudes of
2-dimensional string theory in flat space, linear dilaton background.
For the generic case, when the fluctuations of the fermi surface on
the two sides of the potential are different, we get additional
contributions to the tachyon scattering amplitudes.  By comparing with
early-time bulk scattering amplitudes of tachyon in the effective
tachyon-graviton-dilaton field theory, we are able to identify two of
the leading additional contributions as due to the presence of
perturbations of the background tachyon and the metric.  Roughly
speaking, the string theory tachyon turns out to be the leg-pole
transform of the `average' ($\phi(\tau)$) of the fluid fluctuations on
the two sides of the potential, while the `difference' $\Delta(\tau)$
is now an additional variable in the problem (see Sec. 4 for details).
Both the extra tachyon background and the metric perturbation vanish
if $\Delta(\tau) =0$. The metric pertubation to the leading order
has the form of a linearised black hole metric of mass $M = (1/4 \pi)
\int d\tau\; \Delta^2(\tau)$, which is precisely the extra energy of
the fermi fluid contributed by $\Delta(\tau)$. Besides the contribution
attributable to additional tachyon and metric backgrounds
\footnote{We should emphasize that any additional tachyon background
of course causes curvature and therefore a perturbation to the
metric(back reaction); it is easy to excite {\sl such} metric
perturbations in the conventional framework (ignoring fluid
fluctuations on the other side) by simply declaring one of the pulses
in a scattering experiment as `background'. However, such a metric
perturbation would not constitute independent moduli. On the other
hand, the metric perturbation that we find from the scattering
amplitudes is over and above the back reaction (the latter is also
present in our model, but is of order $\Delta^4$ and is therefore
subleading and distictly identifiable). In the language of string
field theory, in our framework we can tune vacuum expectation values of
metric and tachyon independently.}  there are an infinite series of
subleading contributions to the bulk scattering at early times in
presence of non-zero value of $\Delta(\tau)$. Presumably these reflect
the presence of perturbations in the background values of the other
higher tensor fields of 2-dimensional string theory corresponding to
the higher discrete states \cite{five}.

The plan of this paper is as follows.  To make this paper somewhat
self-contained, in the next section we will review the phase space
approach \cite{six,seven} to the double-scaled matrix model and the
collective field parametrization of the fluid fluctuations.  In Sec. 3
we will discuss the leg-pole transform and recall how the known tachyon
string amplitudes in flat space and linear dilaton background are
obtained using this transform in the existing approach which neglects
the other side of the potential.  In Sec. 4 we will include both sides
in the transform and identify the `average' and `difference' variables
mentioned above.  We will see that the flat space physics of the
string theory is obtained in the limit in which the difference
variable vanishes.  In the generic case, additional leading
contributions to early-time bulk scattering will
be identified.  In Sec. 5 we will show that the above results are
identical to the bulk scattering amplitudes coming from the effective
tachyon-graviton-dilaton action in the presence of background tachyon
and metric perturbations.  The parameters of these perturbations will
be identified in terms of the `difference' variable.  In particular,
we will see that the parameter of the background metric perturbation
is precisely the contribution of the `difference' variable to the
total energy of the fermi fluid fluctuations.  Sec. 6 is devoted to
some concluding remarks in which we will point out the implications of
our work for a consistent nonperturbative definition of string theory
as well as for quantized gravity and black-hole physics.

\section{Review of the Phase Space Formulation of the Matrix Model}

Here we will briefly review some aspects of the phase space
formulation of the $c=1$ matrix model that will be relevant to the
discussion in the following sections.

In the double-scaling limit, the $c=1$ matrix model is mapped to a
model of noninteracting, nonrelativistic fermions in an inverted
harmonic oscillator potential \cite{three} in one space dimension.
The single-particle hamiltonian for this model is
\be
h(p,q) = {1 \over 2} (p^2 - q^2)
\l{2.1}
\ee
where $(p,q)$ labels the single-particle phase space of the fermions.
There is a convenient field theoretic description for the
double-scaled model in terms of free nonrelativistic fermions
\cite{eight}.  The fermion field, which we denote by $\psi(q,t)$,
satisfies the equation of motion
\be
i\partial_t \psi(q,t) = - {1 \over 2}(\partial^2_q + q^2) \psi(q,t)
\l{2.2}
\ee
and its conjugate $\psi^\dagger(q,t)$ satisfies the complex conjugate
of eqn. \eq{2.2}.  The ground state of this model is the fermi vacuum
obtained by filling up to the energy level $\mu$ ($< 0$).  The
semiclassical limit is obtained as $|\mu| \rightarrow \infty$, and in
this limit the fermi surface is described by the hyperbola
\be
{1 \over 2} (p^2 - q^2) = \mu = - |\mu|
\l{2.3}
\ee

The basic building block for this work will be the phase space density
of fermions, which we denote by ${\cal U}(p,q,t)$.  In terms of the
fermi field $\psi(q,t)$ it is defined as
\be
{\cal U}(p,q,t) \equiv \int^{+\infty}_{-\infty} d\lambda \
e^{-ip\lambda} \ \psi^\dagger\left(q - {\lambda \over 2}, t\right)
\psi\left(q + {\lambda \over 2}, t\right),
\l{2.4}
\ee
and it satisfies the equation of motion
\be
\left(\partial_t + p \partial_q + q\partial_p\right) {\cal U}(p,q,t) =
0.
\l{2.5}
\ee
This equation can be obtained using eqn. \eq{2.2}, or by directly
using the hamiltonian
\be
H = \int {dp \ dq \over 2\pi} \ h(p,q) {\cal U}(p,q,t)
\l{2.6}
\ee
and the equal-time commutation relation for the phase space density
${\cal U}(p,q,t)$, which follows from its definition, eqn. \eq{2.4},
in terms of the underlying fermions:
\beq
[{\cal U}(p,q,t),{\cal U}(p',q',t)] &=& -4 \int {dp^{\prime\prime}
dq^{\prime\prime} \over 2\pi} {\cal
U}(p^{\prime\prime},q^{\prime\prime},t) \nonumber \\[2mm]
& & [\exp 2i\{p(q' - q^{\prime\prime}) + p'(q^{\prime\prime}
- q) \nonumber \\[2mm]
& & \hspace{1.3cm} + p^{\prime\prime} (q - q')\} - c.c.]
\l{2.7}
\eeq

Equation \eq{2.7} is also a version of the large symmetry algebra,
$W_\infty$ \cite{nine}, which is a symmetry of the matrix
model \cite{ten}.  The more standard version of the generators of this
symmetry algebra is the following:
\be
W_{mn} = e^{-(m-n)t} \int {dp \ dq \over 2\pi} (-p-q)^m (p-q)^n {\cal
U}(p,q,t),
\l{2.8}
\ee
where $m,n \ \geq \ 0$.  One can easily check, using eqn. \eq{2.5},
that $W_{mn}$ are conserved.  They satisfy the classical algebra
\be
\left\{W_{mn}, W_{m'n'}\right\}_{P.B.} = 2(m'n - mn')
W_{m+m'-1,n+n'-1}.
\l{2.9}
\ee
The quantum version of this is more complicated, but can be computed
using eqn. \eq{2.7}.

The above phase space density formalism was first introduced in the
present context in \cite{six}, and using this variable a bosonization of
the double-scaled matrix model was carried out in \cite{six,seven}.  A
crucial ingredient in that bosonization is a quadratic constraint
satisfied by ${\cal U}(p,q,t)$ \cite{six}.  In the semiclassical limit
this quantum constraint reduces to the simpler equation
\be
{\cal U}^2 (p,q,t) = {\cal U}(p,q,t).
\l{2.10}
\ee
Moreover, one also has the constraint of fixed fermion number, which
implies that fluctuations of the fermi surface satisfy
\be
\int {dp \ dq \over 2\pi} \delta {\cal U} (p,q,t) = 0, \ \ \ \delta
{\cal U}(p,q,t) = {\cal U}(p,q,t)-{\cal U}_0 (p,q),
\l{2.11}
\ee
where ${\cal U}_0 (p,q)$ describes the filled fermi vacuum.  In this
way we recover the Thomas-Fermi limit of an incompressible fermi
fluid.  The dynamics of the fluctuations $\delta {\cal U}(p,q,t)$,
which satisfy eqn. \eq{2.11} and another constraint because of eqn.
\eq{2.10}, resides only in the boundary of the fermi fluid (in the
semiclassical limit that we are considering here) \cite{eleven}.

Although one need not ever use an explicit parametrization of $\delta
{\cal U} (p,q,t)$, it will, nevertheless, be useful at times to do so.
For this reason we summarize in the following some relevant aspects of
the `quadratic profile' \cite{eleven} or `collective field'
\cite{twelve} parametrization of the fluctuations $\delta {\cal U}
(p,q,t)$.

In the semiclassical limit, the fermi vacuum is described by the
density
\be
{\cal U}_0 (p,q) = \theta\left(P^0_+ (q) - p\right) \theta\left(p -
P^0_- (q)\right),
\l{2.12}
\ee
where
\be
P^0_\pm (q) \equiv \pm P_0 (q) = \pm \sqrt{q^2 + 2\mu}
\l{2.13}
\ee
satisfy eqn. \eq{2.3}, which describes the fermi surface hyperbola.
The quadratic profile or collective field description corresponds to a
description of small ripples on the fermi surface by a density of the
form
\be
{\cal U} (p,q,t) = \theta \left(P_+ (q,t) - p\right) \theta\left(p -
P_- (q,t)\right)
\l{2.14}
\ee
Substituting this in eqn. \eq{2.5}, we get the equation of motion of
$P_\pm$:
\be
\partial_t P_\pm (q,t) = {1 \over 2} \partial_q \left(q^2 - P^2_\pm
(q,t)\right).
\l{2.15}
\ee
This equation is clearly solved by the fermi vacuum, eqns. \eq{2.12}
and \eq{2.13}.  Fluctuations around this ground state,
\be
P_\pm (q,t) - P^0_\pm (q) \equiv \eta_\pm (q,t)
\l{2.16}
\ee
satisfy the equations of motion
\be
\partial_t \eta_\pm(q,t) = \mp \partial_q \left[P_0 (q) \eta_\pm (q,t)
\pm {1 \over 2} \eta^2_\pm (q,t)\right].
\l{2.17}
\ee

If the fluctuations are so small that they never cross the asymptotes,
$p = \pm q$, of the hyperbola defined by eqn. \eq{2.3}, then one can
rewrite eqns. \eq{2.17} in a form that exhibits the presence of a
massless particle.  This is done by introducing the time-of-flight
variable $\tau$, defined by
\be
q = \mp |2\mu|^{1 \over 2} \cosh \tau, \ \ \ 0 \leq \tau \ < \ \infty,
\l{2.18}
\ee
where the $-$ve sign is appropriate for the left side $(q \ < \ 0)$ of
the potential and the +ve sign for the right side $(q \ > \ 0)$.

We now introduce the new variables $\bar\eta^\alpha_\pm (\tau,t)$,
$\alpha = 1,2$, defined by
\be
\bar\eta^1_\pm (\tau,t) \equiv P_0 (q(\tau)) \eta_\pm (q,t), \ \ \ q \ < \
0
\l{2.19}
\ee
\be
\bar\eta^2_\pm (\tau,t) \equiv -P_0(q(\tau)) \eta_\mp (q,t), \ \ \ q \
> \ 0
\l{2.20}
\ee
The variables $\bar\eta^\alpha_\pm (\tau,t)$ describe a generic
small-field fluctuation on the two sides of the potential.  They
satisfy the equations of motion
\be
(\partial_t \mp \partial_\tau) \bar\eta^\alpha_\pm (\tau,t) =
\partial_\tau \left[(\bar\eta^\alpha_\pm
(\tau,t))^2/2P^2_0(q(\tau))\right].
\l{2.21}
\ee
Furthermore, one can also deduce the commutation relations
\beq
\left[\bar\eta^\alpha_\pm (\tau,t),\bar\eta^\beta_\pm (\tau,t)\right] &=& \pm
2\pi i \delta^{\alpha\beta} \partial_\tau \delta(\tau - \tau'),
\nonumber \\[2mm]
\left[\bar\eta^\alpha_+ (\tau,t), \bar\eta^\beta_- (\tau,t)\right] &=&
0,
\label{2.22}
\eeq
since we know the hamiltonian for the fluctuations
\beq
H_{fluc.} &=& \int {dp \ dq \over 2\pi} h(p,q) \delta {\cal U}(p,q,t)
\nonumber \\[2mm]
&=& {1 \over 4\pi} \int^\infty_0 d\tau \sum_{\alpha=1,2}
\Bigg[(\bar\eta^\alpha_+ (\tau,t))^2 + (\bar\eta^\alpha_- (\tau,t))^2
+ {1 \over 3P^2_0 (q(\tau))} \nonumber \\[2mm]
& & \hspace{2.7cm}\left\{\left(\bar\eta^\alpha_+ (\tau,t)\right)^3 -
\left(\bar\eta^\alpha_- (\tau,t)\right)^3\right\}\Bigg].
\l{2.23}
\eeq
Finally, the `fixed area' (i.e. fixed fermion number) constraint, eqn.
\eq{2.11}, reads now
\be
\int^\infty_0 d\tau \sum_{\alpha=1,2} \left[\bar\eta^\alpha_+ (\tau,t)
- \bar\eta^\alpha_- (\tau,t)\right] = 0.
\l{2.24}
\ee

In the semiclassical limit, for fluctuations of the fermi surface that
are entirely confined within the asymptotes, eqns. \eq{2.19} --
\eq{2.23} define a set of two decoupled massless scalar fields.
Moreover, the constraint eqn. \eq{2.24} is also satisfied if it is
satisfied for each individual value of $\alpha$.  Therefore, we may
consistently ignore fluctuations on one side of the potential within
the framework of the matrix model.  The mapping to string theory is,
however, a different story, as discussed in the previous section and
as we shall see in detail in Sec. 4.

Before closing this section we mention that in terms of the incoming
fields,
\be
\bar\eta^\alpha_{+in} (t + \tau) \equiv  {\atop {Lt. \atop t \rightarrow
-\infty}} \bar\eta^\alpha_+ (\tau,t),
\l{2.25}
\ee
the hamiltonian, $H_{fluc.}$, and the constraint in eqn. \eq{2.24}
have the following simple expressions
\be
H_{fluc.} = {1 \over 4\pi} \int^{+\infty}_{-\infty} d\tau
\sum_{\alpha=1,2} \left(\bar\eta^\alpha_{+in} (\tau)\right)^2,
\l{2.26}
\ee
\be
\int^{+\infty}_{-\infty} d\tau \sum_{\alpha=1,2} \bar\eta^\alpha_{+in}
(\tau) = 0.
\l{2.27}
\ee
We will use these expressions in Sec. 4.

\section{The Leg-pole Transformation to String Theory}

In this section we will discuss the leg-pole transformation of the
matrix model to string theory within the framework that completely
ignores one side of the potential for fluctuations which are entirely
confined within the asymptotes.  This will set the stage for the
considerations of the next section.

It has been known for some time now that the tree-level scattering
amplitudes for the matrix model scalar (discussed in the last section)
do not exactly coincide with the tree-level scattering amplitudes for
the tachyon in 2-dimensional string theory \cite{one}.  The difference
can be understood in terms of a wavefunction renormalization and is a
simple momentum-dependent phase factor for real momenta.  In
coordinate space this renormalization factor relates the Hilbert space
of the matrix model to that of the string theory by a nonlocal
transform of the states \cite{two}.  Denoting the tachyon field of
2-dimensional string theory by $\T(x,t)$, $(x,t)$ being space-time
labels, this relationship can be expressed as
\be
\T_{in} (x^+) = \int^{+\infty}_{-\infty} d\tau \ f\left(\left|{\mu
\over 2}\right|^{1 \over 2} e^{\tau - x^+}\right)\bar\eta_{+in}
(\tau),
\l{3.1}
\ee
\be
\T_{out} (x^-) = -\int^{+\infty}_{-\infty} d\tau \ f\left(\left|{\mu
\over 2}\right|^{1 \over 2} e^{-\tau + x^-}\right)\bar\eta_{-out}
(\tau),
\l{3.2}
\ee
where $x^\pm \equiv (t \pm x)$ and the `$in$' and `$out$' refer, as
usual, to the asymptotic fields obtained in the limits $t \rightarrow
-\infty$ and $t \rightarrow +\infty$ respectively.  In both cases $x$
is taken to be large positive, keeping respectively $x^+$ and $x^-$
fixed.  The function $f$ is given by
\be
f(\sigma) \equiv {1 \over 2\sqrt{\pi}} J_0\left(2\left({2 \over
\pi}\right)^{1 \over 8} \sqrt{\sigma}\right), \ \ \ \sigma \geq 0
\l{3.3}
\ee
where $J_0$ is the standard Bessel function of order zero
\cite{thirteen}.

That the above nonlocal transformation is essential for extracting the
space-time physics of 2-dimensional string theory from the matrix
model has only recently become clear \cite{two,four}.  It contains all
of the space-time gravitational physics of the string theory, which is
absent in the matrix model.  Moreover, a detailed operator mapping
from the latter to the former emerges only after the above nonlocal,
and in general a nonlinear (in matrix model variables),
transformation.

In general, the transform, which is valid even away from the
asymptotic space-time region of eqns. \eq{3.1} and \eq{3.2}, is a
complicated nonlinear and nonlocal combination of the matrix model
scalar \cite{fourteen}.  The general mapping may be written as
\beq
\T(x,t) &\equiv& \int dp \ dq \ G_1(x;p,q) \delta {\cal U}(p,q,t)
\nonumber \\[2mm] & & + {1\over2} \int dp \ dq \int dp' \ dq' \
G_2(x;p,q;p',q') \delta {\cal U}(p,q,t) \delta {\cal U}(p',q',t)
\nonumber \\[2mm] & & + \cdots
\l{3.4}
\eeq
where the dots stand for terms of higher order in $\delta {\cal U}$.
The kernels $G_1$, $G_2$, etc. that are necessary for $\T(x,t)$ to
satisfy the tachyon $\beta$-function equation of string theory and the
canonical commutation relations of a scalar field in flat space,
linear dilaton background are known \cite{fourteen} (upto corrections
of order $xe^{-4\,x}$).

For the purposes of computing tree level tachyon scattering amplitudes
only the asymptotic form of the kernels is relevant, since the
corrections drop out at large +ve values of $x$, as argued in
\cite{two,fourteen}.
Since in this section we are ignoring one side of the potential, we
shall assume that $\delta {\cal U} (p,q,t) = 0$ for $q \ > \ 0$ and
restrict our attention to only -ve values of $q$.
The asymptotic form of the mapping is, then,  given by
\be
\T(x,t) = \int dp \ dq \ f(-qe^{-x}) \delta {\cal U}(p,q,t) +
O(xe^{-2x}).
\l{3.5}
\ee
Although this equation reproduces eqns. \eq{3.1} and \eq{3.2} for
quadratic profiles, in this form it is valid for any arbitrary
fluctuation, not necessarily of the quadratic profile form.

The scattering
amplitudes may now be computed by shifting the entire $t$-dependence
of $\delta {\cal U}(p,q,t)$ into the argument of the function $f$,
using the equation of motion
\be
(\partial_t + p\partial_q + q\partial_p) \delta {\cal U} (p,q,t) = 0.
\l{3.6}
\ee
According to this equation $\delta {\cal U}(p,q,t) = \delta {\cal U}
(p',q',t')$, where $(p' \pm q') e^{\mp t'} = (p \pm q)^{\mp t}$. We
use this in the integral in eqn. \eq{3.5} to change variables from
$(p,q)$ to $(p',q')$, with $t$ and $t'$ as fixed parameters for the
purposes of this change of variables.  Under this change of
variables the measure $(dp\ dq)$ and the fermi surface,
defined by eqn. \eq{2.3}, are invariant.
Therefore, making this change of variables in eqn. \eq{3.5}, we get
\be
\T(x,t) = \int dp \ dq \ f(-Q(t)e^{-x}) \delta {\cal U}(p,q,t_0) +
O(xe^{-2x}),
\l{3.7}
\ee
where
\be
Q(t) \equiv q \cosh (t - t_0) + p \sinh(t - t_0)
\l{3.8}
\ee
The right hand side of eqn. \eq{3.7} can be proved to be independent
of the parameter $t_0$, using eqn. \eq{3.6}, and shows that the fermi
fluid fluctuation enters eqn. \eq{3.5} only as a boundary condition.

Let us now assume that $\delta {\cal U}(p,q,t_0)$ is of the quadratic
profile form and let us further use the $t_0$-independence of the
r.h.s. of eqn. \eq{3.7} to take $t_0 \rightarrow -\infty$.  Under
these assumptions $\delta {\cal U} (p,q,t_0)$ is parametrized by the
single field $\bar\eta^1_{+in} (\tau) \equiv \eta(\tau)$ discussed in
Sec. 2 (since $\delta {\cal U} = 0$ for $q > 0$).  Using this and the
formalism developed in Sec. 2 in eqn. \eq{3.7} in the limit $t_0
\rightarrow -\infty$, we get
\be
\T(x,t) = \int^{+\infty}_{-\infty} d\tau \int^{\eta(\tau)}_0 d\epsilon
\ f\left(\left|{\mu \over 2}\right|^{1\over2} \left[e^{\tau-x^+} +
\left(1 - {\epsilon \over |\mu|}\right) e^{-\tau+x^-}\right]\right) +
O(xe^{-2x}).
\l{3.9}
\ee
It is now trivial to write down expressions for the `$in$' and `$out$'
fields from this equation.  We get
\be
\T_{in} (x^+) = \int^{+\infty}_{-\infty} d\tau \ \eta(\tau)
f\left(\left|{\mu \over 2}\right|^{1\over2} e^{\tau - x^+}\right),
\l{3.10}
\ee
\be
\T_{out}(x^-) = \int^{+\infty}_{-\infty} d\tau \int^{\eta(\tau)}_0
d\epsilon \ f\left(\left|{\mu \over 2}\right|^{1\over2} \left(1 -
{\epsilon \over |\mu|}\right) e^{-\tau+x^-}\right).
\l{3.11}
\ee
The tree level scattering amplitudes may now easily be obtained by
inverting eqn. \eq{3.10} for $\eta(\tau)$,
\be
\eta(\tau) = -4\pi \int^{+\infty}_{-\infty} dx^+ \ \T_{in} (x^+)
\partial^2_\tau f\left(\left|{\mu \over 2}\right|^{1\over2} e^{\tau
-x^+}\right),
\l{3.12}
\ee
and substituting in $\T_{out}$ after expanding it in a power series in
$\eta(\tau)$.  It is a simple exercise to check that the correct
string tachyon scattering amplitudes in flat space, linear dilaton
background are obtained in this way.

\section{Transforming fluctuations on both sides of the potential}

In the previous section we applied the leg-pole transformation to
fermi surface fluctuations only on one side of the potential, ignoring
the other side.  As we have argued in Sec. 1, this procedure is
inconsistent with the space-time gravitational physics that one wants
to extract from the matrix model, unless one considers a modified
potential in which one side is absent right from the start.  Here we
will consider the case in which the potential is unmodified.  In that
case, therefore, we must include fluctuations on both sides for a
consistent space-time gravitational physics to emerge after the
transformation.

Let us, then, consider the generic case of fluctuations on both sides
of the potential (Fig. 3).  As before, we will assume that the
fluctuations are small and entirely confined within the asymptotes.

We must now decide as to what generalization of the leg-pole transform
considered in the previous section (in which $q$ was always -ve)
should be taken for $q$ +ve.  The symmetry between the two sides of
the potential suggests that we try the following symmetrical leg-pole
transform
\be
\T (x,t) = {1 \over \sqrt{2}} \int dp dq \ f(2^{1\over4} |q| e^{-x})
\delta {\cal U}(p,q,t) + O(xe^{-2x}),
\l{4.1}
\ee
where now $\delta u \neq 0$ for both $q < 0$ and $q > 0$.  The overall
factor of $\displaystyle{1 \over \sqrt{2}}$ and the extra factor of
$2^{1\over4}$ in the argument of the function $f$ relative to the
expression in eqn. \eq{3.5}
have been put there for convenience only.
The appearance of $|q|$ in the argument of $f$ is suggested by the
symmetry of the potential.

We would now like to ask the question as to whether the field $\T
(x,t)$ defined in eqn. \eq{4.1} could reproduce string theory
tachyon scattering amplitudes.  To answer this question we need the
analogues of eqns. \eq{3.10} and \eq{3.11} of the previous
section in the present case.  To derive these we proceed exactly as
before.  The analogue of eqn. \eq{3.7} is
\be
\T (x,t) = {1\over\sqrt{2}} \int dp dq \ f(2^{1\over4} |Q(t)| e^{-x})
\delta {\cal U}(p,q,t_0) + O(x\bar e^{2x}),
\l{4.2}
\ee
where $Q(t)$ is given, as before, by eqn. \eq{3.8}.  Using the
$t_0$-independence of eqn. \eq{4.2} to take the limit $t_0
\rightarrow -\infty$ and assuming once again that $\delta {\cal U}(p,q,t_0)$
has the quadratic profile form on both sides of the potential, we get
\beq
\T (x,t) &=& {1\over2} \int^{+\infty}_{-\infty} d\tau \sum_{\alpha=1,2}
\int^{\sqrt{2} \bar\eta^\alpha_{+in} (\tau)}_0 d\epsilon \nonumber \\[2mm]
& & f\left(\left|{\mu' \over 2}\right|^{1\over2} \left[e^{\tau - x^+} +
\left(1 - {\epsilon \over |\mu'|}\right)e^{-\tau+x^-}\right]\right) +
O(x\bar e^{2x})
\l{4.3}
\eeq
where $x^\pm \equiv t \pm x$ as before and $\mu' \equiv \sqrt{2} \mu$.
We can now write down expressions for the `$in$' and `$out$' fields.
They are
\be
\T_{in} (x^+) = \int^{+\infty}_{-\infty} d\tau \left({1 \over
\sqrt{2}} \sum_{\alpha=1,2} \bar\eta^\alpha_{+in} (\tau)\right)
f\left(\left|{\mu' \over 2}\right|^{1\over2} e^{\tau-x^+}\right)
\l{4.4}
\ee
and
\be
\T_{out} (x^-) = {1\over2} \int^{+\infty}_{-\infty} d\tau
\sum_{\alpha=1,2} \int^{\sqrt{2}\bar\eta^\alpha_{+in} (\tau)}_0
d\epsilon f\left(\left|{\mu' \over 2}\right|^{1\over2} \left(1 -
{\epsilon \over |\mu'|}\right) e^{-\tau+x^-}\right).
\l{4.5}
\ee

The form of eqns. \eq{4.4} and \eq{4.5} suggests that the matrix model
variable whose leg-pole transform has the potential of reproducing the
string theory physics be identified with the combination appearing in
eqn. \eq{4.4}, namely,
\be
{1 \over \sqrt{2}} \sum_{\alpha=1,2} \bar\eta^\alpha_{+in} (\tau)
\equiv \phi (\tau).
\l{4.6}
\ee
However, this immediately poses a problem, since the fields
$\bar\eta^1_{+in} (\tau)$ and $\bar\eta^2_{+in} (\tau)$ appear
individually in eqn. \eq{4.5} and not as the sum $\phi(\tau)$.  To see
the implication of this, let us introduce the other combination of
$\bar\eta^1_{+in} (\tau)$ and $\bar\eta^2_{+in} (\tau)$, namely,
\be
{1 \over \sqrt{2}} \left(\bar\eta^1_{+in} (\tau) - \bar\eta^2_{+in}
(\tau) \right) \equiv \Delta (\tau).
\l{4.7}
\ee
In terms of the variables $\phi(\tau)$ and $\Delta (\tau)$, the
Hamiltonian, eqn. \eq{2.26}, and the `fixed area' constraint, eqn.
\eq{2.27}, are
\be
H_{fluc.} = {1 \over 4\pi} \int^{+\infty}_{-\infty} d\tau \left(\phi^2
(\tau) + \Delta^2 (\tau)\right)
\l{4.8}
\ee
and
\be
\int^{+\infty}_{-\infty} d\tau \ \phi(\tau) = 0.
\l{4.9}
\ee
Note that there is no constraint on $\Delta (\tau)$.

Equations \eq{4.4} and \eq{4.5} may now be recast into the following
form
\be
\T_{in} (x^+) = \int^{+\infty}_{-\infty} d\tau \ \phi(\tau)
f\left(\left|{\mu' \over 2}\right|^{1\over2} e^{\tau-x^+}\right),
\l{4.10}
\ee
\beq
\T_{out} (x^-) &=& {1\over2} \int^{+\infty}_{-\infty} d\tau
\Bigg[\int^{\phi(\tau) +\Delta(\tau)}_0 d\epsilon f\left(\left|{\mu'
\over 2}\right|^{1\over2} \left(1 - {\epsilon \over |\mu'|}\right)
e^{-\tau+x^-} \right) \nonumber \\[2mm]
& & + \int^{\phi(\tau) - \Delta(\tau)}_0 d\epsilon
f\left(\left|{\mu' \over 2}\right|^{1\over2} \left(1 - {\epsilon \over
|\mu'|} \right) e^{-\tau + x^-}\right)\Bigg].
\l{4.11}
\eeq
Comparing these two equations with eqns. \eq{3.10} and \eq{3.11} of
the previous section, we see immediately that in the simple case of
$\Delta (\tau) = 0$ we recover the tree level tachyon scattering
amplitudes of string theory in the background of flat space, linear
dilaton (except for a rescaling of the string coupling, $\mu
\rightarrow \sqrt{2} \mu \equiv \mu'$).  Thus, \underbar{in the
present framework these results of flat background emerge only when} \\
\underbar{the fluctuations of the fermi surface on the two sides
of the potential are identical}.

Equations \eq{4.10} and \eq{4.11}, of course, describe the generic
situation of different fluctuations on the two sides and it is natural
to ask whether these equations have a sensible space-time
interpretation for $\Delta (\tau) \neq 0$.  To investigate this
question, let us for now assume that $\Delta (\tau)$ is small, so that
we may Taylor expand eqn. \eq{4.11} around $\Delta (\tau) = 0$.
Retaining only upto the first nontrivial term in $\Delta (\tau)$, we
get
\be
\T_{out} (x^-) = \T^{(0)}_{out} (x^-) + \T^{(1)}_{out} (x^-) + \cdots
\l{4.12}
\ee
where
\be
\T^{(0)}_{out} (x^-) \equiv \int^{+\infty}_{-\infty} d\tau
\int^{\phi(\tau)}_0 d\epsilon \ f\left(\left|{\mu' \over
2}\right|^{1\over2} \left(1 - {\epsilon \over |\mu'|}\right) e^{-\tau
+ x^-}\right),
\l{4.13}
\ee
\be
\T^{(1)}_{out} (x^-) \equiv - {1 \over 2|2\mu'|^{1\over2}}
\int^{+\infty}_{-\infty} d\tau \ \Delta^2(\tau) \ e^{-\tau+x^-} \
f'\left(\left|{\mu' \over 2}\right|^{1 \over 2} \left(1 - {\phi(\tau)
\over |\mu'|}\right) e^{-\tau + x^-}\right)
\l{4.14}
\ee
and the dots in eqn. \eq{4.12} indicate $O(\Delta^4)$ terms.  The
`prime' on the function $f$ in eqn. \eq{4.14} denotes a derivative
with respect to its argument.  The first term in the expansion is the
one that gives the flat space results.  It is the second term that we
would now like to focus on.

Notice that the second term in \eq{4.12} is not zero
\underbar{even when $\phi(\tau)$ is set to zero}.  Since we are
already committed to interpreting the transform of $\phi (\tau)$ as
the string theory tachyon, we are forced to interpret this extra
contribution at $\phi = 0$ to $\T_{out}$ coming from the
second term in eqn. \eq{4.12},
as a new contribution to the tachyon background, dynamically generated
by a nonzero value of $\Delta$.  For such an interpretation to be
self-consistent, however, we should find additional contributions to
the amplitudes for the bulk scattering of tachyons coming from this
extra contribution to the tachyon background.  In eqn. \eq{4.12} there
are indeed extra contributions to tachyon scattering amplitudes, since
the additional $\Delta$-dependent term depends on $\phi$ also.  It is
these extra contributions to the tachyon scattering amplitudes that we
would now like to study in detail.

Let us Taylor expand \eq{4.14} in powers of $\phi$ and retain terms
upto linear order only.  This is because we want to focus here only on
$1 \rightarrow 1$ scattering off the backgrounds.

We get,
\beq
\T^{(1)}_{out} (x^-) &=& - {1 \over 2|2\mu'|^{1\over2}}
\int^{+\infty}_{-\infty} d\tau \ \Delta^2(\tau) e^{-\tau+x^-}
\nonumber \\[2mm] & & \left[f'\left(\left|{\mu' \over
2}\right|^{1\over2} e^{-\tau+x^-}\right) - {\phi(\tau) e^{-\tau+x^-}
\over |2\mu'|^{1\over2}} f^{\prime\prime} \left(\left|{\mu' \over
2}\right|^{1\over2} e^{-\tau+x^-}\right)\right] \nonumber \\[2mm]
& & + O(\phi^2).
\l{4.15}
\eeq
In what follows we will focus on the $1 \rightarrow 1$ bulk scattering
amplitudes at early times $(x^- \rightarrow -\infty)$.  We will also
assume that the incoming tachyon wavefunction $\T_{in} (x^+)$ is a
sufficiently localized (e.g. gaussian) wavepacket concentrated at a
very large positive value of $x^+$.  This is because it is only under
this condition that the background created by a nonzero value of
$\Delta$ will have a chance to develop before the tachyon scatters off
it.  Finally, we will also assume that $\Delta (\tau)$ is sufficiently
localized for integrals like $\displaystyle \int^{+\infty}_{-\infty}
d\tau \ \Delta^2(\tau) e^{-n\tau}$ to be finite.

Now, at early times $(x^- \rightarrow -\infty)$, the first two leading
contributions to $\T^{(1)}_{out}$ are given by
\beq
\T^{(1)}_{out} (x^-) \ {\buildrel x^- \rightarrow -\infty \over \sim} &-&
{f'(0) \over 2|2\mu'|^{1\over2}} e^{x^-} \int^{+\infty}_{-\infty}
d\tau \ \Delta^2(\tau) e^{-\tau} \nonumber \\[2mm]
&+& {f^{\prime\prime} (0) \over 4|\mu'|} e^{2x^-}
\int^{+\infty}_{-\infty} d\tau \ \Delta^2 (\tau) e^{-2\tau} \phi(\tau)
\nonumber \\[2mm] &+& O(\phi^2)
\l{4.16}
\eeq
It is the first term in eqn. \eq{4.16} that we would like to interpret
as a new contribution to the tachyon background and it is the second
term in eqn. \eq{4.16} that we would like to show contains the
amplitude for scattering off this background.

To cast the second term in eqn. \eq{4.16} into a form from which we can
simply read off the scattering amplitude, we need to invert eqn.
\eq{4.10}, using eqn. \eq{4.9}, to express $\phi(\tau)$ in terms of
$\T_{in} (x^+)$.  The expression for $\phi (\tau)$ obtained in this
way is identical to eqn. \eq{3.12},
\be
\phi (\tau) = -4\pi \int^{+\infty}_{-\infty} dx^+ \ \T_{in} (x^+)
\partial^2_\tau \ f\left(\left|{\mu' \over 2}\right|^{1\over2}
e^{\tau-x^+}\right)
\l{4.17a}
\ee
Using this, we get
\beq
\int^{+\infty}_{-\infty} &d\tau& \Delta^2 (\tau) e^{-2\tau}\phi(\tau)
\nonumber \\[2mm] &=& -4\pi \int^{+\infty}_{-\infty} dx^+ \T_{in} (x^+)
\int^{+\infty}_{-\infty} d\tau \ \Delta^2(\tau) e^{-2\tau}
\partial^2_\tau f\left(\left|{\mu' \over 2}\right|^{1\over2}
e^{\tau-x^+}\right) \nonumber \\[2mm]
&=& -4\pi f'(0) \left|{\mu' \over 2}\right|^{1\over2}
\int^{+\infty}_{-\infty} d\tau \ \Delta^2(\tau) e^{-\tau}
\int^{+\infty}_{-\infty}  dx^+ \ e^{-x^+} \T_{in} (x^+) \nonumber
\\[2mm]
& & - 4\pi f^{\prime\prime} (0) |\mu'| \int^{+\infty}_{-\infty} d\tau \
\Delta^2 (\tau) \int^{+\infty}_{-\infty} dx^+ \ e^{-2x^+} \T_{in}
(x^+) \nonumber \\[2mm]
& & +\cdots \nonumber
\eeq
where in writing the last step we have used the assumption that
$\T_{in} (x^+)$ is localized at a large +ve value of $x^+$ to expand
the function $f$ in a Taylor series, and retained only the first two
terms.  Putting all this in \eq{4.16} we finally get

\newpage

\beq
\T^{(1)}_{out} (x^-) \ {\buildrel x^- \rightarrow -\infty \over \sim}
&-& {f'(0) \over 2|2\mu'|^{1\over2}} e^{x^-} \int^{+\infty}_{-\infty}
d\tau \ \Delta^2(\tau) e^{-\tau} \nonumber \\[2mm]
&-& {f'(0) \over 2\sqrt{2}|2\mu'|^{1\over2}} e^{2x^-} \int^{+\infty}_{-\infty}
d\tau \ \Delta^2(\tau) e^{-\tau} \nonumber \\[2mm]
& & \hspace{1cm} \times \int^{+\infty}_{-\infty} dx^+ \ e^{-x^+}
\T_{in} (x^+) \nonumber \\[2mm]
&-& {1 \over 8\pi} e^{2x^-} \int^{+\infty}_{-\infty} d\tau \ \Delta^2
(\tau) \int^{+\infty}_{-\infty} dx^+ \ e^{-2x^+} \T_{in} (x^+)
\l{4.17}
\eeq
In the above we have used that $f^{\prime\prime} (0) = \displaystyle{1
\over 2\sqrt{2} \pi}$.

In the next section we will show that the second term in eqn.
\eq{4.17} arises as a result of tachyon scattering in the presence of
a tachyon background given by the first term in eqn. \eq{4.17}.  We
will also show that the last term in eqn. \eq{4.17} arises from
tachyon scattering off
a background metric, perturbed around flat space, and of the form
given by the line element
\be
(ds)^2 = (1 - Me^{-4x})dt^2 - (1 + Me^{-4x})dx^2,
\l{4.18}
\ee
where the parameter $M$ of perturbation is given by
\be
M \equiv {1 \over 4\pi} \int^{+\infty}_{-\infty} d\tau \
\Delta^2(\tau).
\l{4.19}
\ee
This is identical to the contribution of $\Delta(\tau)$ to the total
energy of the fluctuations, $H_{fluc.}$, eqn. \eq{4.8}.

\section{Effective Low-energy String Theory}

Let us now verify in detail the claims made above by a calculation in
the known tachyon-dilaton-graviton effective field theory limit of
2-dimensional string theory \cite{fifteen}.

The field equations of this effective field theory are
\beq
R_{\mu\nu} + 2\nabla_\mu \nabla_\nu \Phi - 2\partial_\mu T
\partial_\nu T &=& 0, \nonumber\\[2mm]
R + 4\nabla^2\Phi - 4(\nabla\Phi)^2 - 2(\nabla T)^2 - 8T^2 &=& 16,
\nonumber \\[2mm] \nabla^2T - 2\nabla\Phi \nabla T - 4T - 2\sqrt{2} T^2 &=&
0.
\l{5.1}
\eeq
We will work in the gauge in which the dilaton is always given by its
linear form (at least in a local neighbourhood), namely,
\[
\Phi = -2x,
\]
and the metric has the diagonal form
\[
g_{\mu\nu} = diag(g_{00},g_{11}).
\]
Perturbing around the flat background
\be
g_{00} = 1 - g_0, \ \ \ g_{11} = -(1 + g_1),
\l{5.2}
\ee
and retaining only upto linear order in tachyon fluctuations and upto
the first nontrivial order in the tachyon backgrounds, we get the
following equations for the metric and tachyon fluctuations:
\beq
\partial_x g_0 + 4g_1 &=& 0, \nonumber \\[2mm]
(\partial_x + 4)g_1 &=& 0, \nonumber \\[2mm]
\partial_t g_1 &=& 0,
\l{5.3}
\eeq
and
\be
\partial_+ \partial_- S = - {1\over4} g_0 \partial^2_t S - {1\over4}
g_1 (\partial^2_x + 4\partial_x - 12)S + \sqrt{2} e^{-2x} S_0 S,
\l{5.4}
\ee
where we have set $T = e^{-2x} S$ and $S_0$ is the tachyon background.

The eqns. \eq{5.3} have the black-hole solution \cite{sixteen}
\be
g_0 = g_1 = \tilde M e^{-4x},
\l{5.5}
\ee
where $\tilde M$ is a constant of integration.  Using this solution in
eqn. \eq{5.4} and integrating it to first order in the background
$S_0$, with the boundary condition
\[
S(x,t) \ {\buildrel t \rightarrow -\infty \over \longrightarrow} \
S_{in} (x^+),
\]
we get
\beq
S_{out} (x^-) &=& {1 \over \sqrt{2}} e^{2x^-} \int^{+\infty}_{-\infty}
dx^+ \ e^{-x^+} \ \tilde S_0(x^-,x^+) S_{in} (x^+) \nonumber \\[2mm]
& & - {\tilde M \over 2} e^{2x^-} \int^{+\infty}_{-\infty} dx^+
e^{-2x^+} S_{in} (x^+).
\l{5.6}
\eeq
In this equation $S_{out} (x^-) = {\atop {Lt. \atop t \rightarrow
+\infty}} S(x,t)$ and
\[
\tilde S_0 (x^-,x^+) \equiv 2e^{-2x^-} \int^{x^-}_{-\infty} du^- \
e^{u^-} \ S_0(u^-,x^+).
\]
For $S_0$ given by the first term in eqn. \eq{4.17}, we get
\[
\tilde S_0 (x^-,x^+) = - {f'(0) \over 2|2\mu'|^{1\over2}}
\int^{+\infty}_{-\infty} d\tau \ \Delta^2 (\tau) e^{-\tau}
\]
We thus see that eqn. \eq{5.6} is identical to the last two terms of
eqn. \eq{4.17}, provided we identify $S$ with $\T$, $S_0$ with the
first term in eqn. \eq{4.17} and $\tilde M$ with $M$ given in eqn.
\eq{4.19}.  This proves the claims made in the previous section.

To end this section we mention that the full contribution (at order
$\Delta^2$, but to all orders in $e^{x^-}$) to what we would like to
identify as a perturbation to the tachyon background, given by the
first term in eqn. \eq{4.15}, is
\be
\T^{bgd.}_{out} (x^-) \equiv - {1 \over 2|2\mu'|^{1\over2}}
\int^{+\infty}_{-\infty} d\tau \ \Delta^2(\tau) e^{-\tau+x^-}
f'\left(\left|{\mu' \over 2}\right|^{1\over2} e^{-\tau+x^-}\right) +
O(\Delta^4)
\l{5.7}
\ee
One can easily check that it is a consistent interpretation of this
contribution to $\T_{out}$, since the term linear in $\phi(\tau)$ in
eqn. \eq{4.15} contains a contribution that equals the amplitude of
the tachyon to scatter off this background, as calculated in the
effective field theory above.

We also mention here that the full contribution to the $1 \rightarrow
1$ background scattering of the tachyon (including all the nonleading
terms in $e^{x^-}$ as well as $e^{-x^+}$ for $x^- \rightarrow -\infty$
and $x^+ \rightarrow +\infty$ respectively), as given by the second
term in eqn. \eq{4.15}, is
\beq
&-& {\pi \over |\mu'|} \int^{+\infty}_{-\infty} dx^+ \ \T_{in} (x^+)
\int^{+\infty}_{-\infty} d\tau \ \Delta^2(\tau) e^{-2\tau + 2x^-}
\nonumber \\[2mm]
& & \hspace{.5cm} \partial^2_\tau f\left(\left|{\mu' \over 2}\right|^{1\over2}
e^{\tau-x^+}\right) f^{\prime\prime} \left(\left|{\mu' \over
2}\right|^{1\over2} e^{-\tau+x^-}\right). \nonumber
\eeq
We have used eqn. \eq{4.17} in arriving at this expression.  Assuming,
as before, that $\T_{in} (x^+)$ is a localized wavepacket concentrated
around a large positive value of $x^+$, and taking the early time
limit, $x^- \rightarrow -\infty$, we can rewrite the above as a double
series in $e^{-x^+}$ and $e^{x^-}$ by expanding both the Bessel
functions around the origin.  The final result can be written in the
compact form
\be
\sum^\infty_{m,n=1} C_{mn} e^{(n+1)x^-} \int^{+\infty}_{-\infty} dx^+
e^{-(m+1)x^+} \T_{in} (x^+)
\l{5.8}
\ee
where
\beq
C_{mn} &\equiv& {(-)^{m+n+1} \over 4\pi} \left|{\mu' \over
\sqrt{2\pi}}\right|^{{m+n \over 2} -1} {1 \over (m!)^2 (n+1)! (n-1)!}
\times \nonumber \\[2mm]
& & \hspace{.5cm} \times \int^{+\infty}_{-\infty} d\tau \
\Delta^2(\tau) e^{(m-n)\tau}
\l{5.9}
\eeq
We have omitted the $m=0$ (arbitrary $n$) term from the sum in eqn.
\eq{5.8} which, as we have already seen, corresponds to a perturbation
of the tachyon background.  The $m=n=1$ term is due to the metric
perturbation while the other terms have a structure that is consistent
with their interpretation as scattering off backgrounds corresponding
to the higher discrete states of 2-dimensional string theory.
Unfortunately we do not have an effective theory to guide us here.
But perhaps one could turn this situation around and use eqn. \eq{5.8}
to learn something about such a theory.  However, we shall not pursue
this interesting problem here any further.

\section{Quantum gravity, Black hole physics and
Nonperturbative String Theory}

Our discussion has so far been classical.  By quantizing $\phi(\tau)$
in the standard manner, and treating $\Delta(\tau)$ as a classical
variable, we can convert our classical statements into statements
about tree level scattering amplitudes in string theory in the
presence of space-time background fields.  As long as the fermi
surface fluctuations are confined within the asymptotes and the string
coupling is small (so that tunnelling amplitudes are supressed), the
two sides of the matrix model potential are decoupled from each other
and so are the fields $\phi(\tau)$ and $\Delta(\tau)$.  It is then
consistent to quantize $\phi(\tau)$ while treating $\Delta (\tau)$ as
a classical variable.  If the string coupling is large, however, there
will be appreciable tunnelling from one side of the potential to the
other.  In that case we can no longer treat the two sides of the
potential as decoupled from each other and so we must quantize the
entire system.  Thus, the backgrounds are dynamical fields and there
are really no parameters in the theory, as is expected in a string
theory.  (Note that even the string coupling, $|\mu|^{-1}$, is not a
parameter in this model since it can be absorbed into the constant
part of the dilaton by shifting $x$.)  In particular, this means that
we must quantize the background geometry.  It is satisfying that this
scenario emerges in the present framework since one does indeed expect
the classical picture of space-time geometry to break down in strong
coupling string theory.  It is to be hoped that the results of this
work, together with the full power of the nonperturbative solution of
the $c=1$ matrix model, will give us a concrete handle on this
question.

The picture of two decoupled sides of the potential breaks down even
for small values of the string coupling, when one is considering
certain special fermi fluid configurations.  These are the
configurations in which part of the fluid crosses the asymptotes
(Figure 2).  This is because for such configurations some of the fluid
eventually goes across to the other side.  In the framework in which
one side of the potential is completely ignored in the leg-pole
transform, such configurations lead to a nonperturbative consistency
problem \cite{seventeen}, basically because some of the fluid is lost
to the other side.  In the framework considered in the present paper,
in which both sides of the potential are always taken into account, no
fluid can ever get lost.  Thus the problem as posed in
\cite{seventeen} does not occur here.  Nevertheless, it does not
follow that there is no problem for such configurations, and we need
to establish the consistency of our framework for such configurations.
This situation is made complicated by the fact that there is a
dynamical rearrangement of the fluid between the two sides, and so it
is not clear that it is consistent to treat $\phi(\tau)$ quantum
mechanically while $\Delta(\tau)$ is treated classically.  It is, in
fact, quite conceivable that such configurations are accompanied by a
change of the background fields and, in particular, of the background
geometry, from the given initial state to some other final state.  If
this is true, then we would have discovered a way of dynamically
changing backgrounds in our framework.  It is possible, then, to
eventually hope to get a handle on the formation and evaporation of a
black-hole in this exactly solvable model.  Although we are far from
realizing this at the moment, we are encouraged by the fact that our
framework contains both the background metric perturbation as well as
some as yet little understood large-field classical situations.

Finally, we would like to end with the following remark.  The presence
of background perturbations corresponding to the discrete states of
2-dimensional string theory, in the framework which retains both sides
of the potential and treats them consistently, is a strong indication
that a consistent nonperturbative definition of 2-dimensional string
theory must include both sides of the potential.  To use the full
nonperturbative power of the matrix model to address some of the basic
questions of string theory and quantum gravity, it is clearly of
utmost importance to discover this nonperturbative formulation of the
string theory in terms of the matrix model.

\vspace{5 ex}
\noindent{\bf Acknowledgement:} S.W. and G.M. would like to
acknowledge the hospitality of the theory division of CERN where
most of this work was done.



\begin{thebibliography}{99}

\bibitem{one} P. Di Francesco and D. Kutasov, Phys. Lett. \und{B261}
(1991) 385; Nucl. Phys. B375 (1992) 119; D.J. Gross and I.R. Klebanov,
Nucl. Phys. \und{B359} (1991) 3; N. Sakai and Y. Tanii, Prog. Theor.
Phys. Suppl. \und{110} (1992) 117; Phys. Lett. \und{B276} (1992) 41;
G. Minic and Z. Yang, Phys. Lett. \und{B274} (1992) 27; D. Lowe, Mod.
Phys. Lett. \und{A7} (1992) 2647.
\bibitem{two} M. Natsuume and J. Polchinski, Nucl. Phys. \und{B424}
(1994) 137.
\bibitem{three} E. Br\'ezin, V.A. Kazakov and A.B. Zamolodchikov, Nucl.
Phys. \und{B333} (1990) 673; P. Ginsparg and J. Zinn-Justin, Phys.
Lett. \und{B240} (1990) 333; G. Parisi, Phys. Lett. \und{B238} (1990)
209; D.J. Gross and N. Miljkovi\'c, Phys. Lett. \und{B238} (1990) 217.
\bibitem{four} J. Polchinski, `What is string theory?', hep-th/9411028.
\bibitem{five} D.J. Gross, I.R. Klebanov and M.J. Newman, Nucl. Phys.
\und{B350} (1990) 621; A.M. Polyakov, Mod. Phys. Lett. \und{A6} (1991)
635.
\bibitem{six} A. Dhar, G. Mandal and S.R. Wadia, Mod. Phys. Lett.
\und{A7} (1992) 3129.
\bibitem{seven} A. Dhar, G. Mandal and S.R. Wadia, Mod. Phys. Lett.
\und{A8} (1993) 3557.
\bibitem{eight} A.M. Sengupta and S.R. Wadia, Int. J. Mod. Phys.
\und{A6} (1991) 1961; D.J. Gross and I. Klebanov, Nucl. Phys.
\und{B352} (1990) 671; G. Moore, Nucl. Phys. \und{B368} (1992) 557.
\bibitem{nine} For a review of various linear and nonlinear higher
spin algebras see, C.N. Pope, L.J. Romans and X. Shen, in Proceedings
of Strings '90, eds. R. Arnowitt, et. al. (World Scientific, 1991)
287; P. Bouwknegt and K. Schoutens, Phys. Rept. \und{223} (1993) 183;
X. Shen, Int. J. Mod. Phys. \und{A7} (1992) 6953.
\bibitem{ten} J. Avan and A. Jevicki, Phys. Lett. \und{B266} (1991)
35; G. Moore and N. Seiberg, Int. J. Mod. Phys. \und{A7} (1992) 2601;
S.R. Das, A. Dhar, G. Mandal and S.R. Wadia, Int. J. Mod.
Phys. \und{A7} (1992) 5165; I. Klebanov and A.M. Polyakov,
Mod. Phys. Lett. \und{A6} (1991) 3273; E. Witten,
Nucl. Phys. \und{B373} (1992) 187; E. Witten and B. Zweibach,
Nucl. Phys. \und{B377} (1992) 55.
\bibitem{eleven} J. Polchinski, Nucl. Phys. \und{B362} (1991) 125;
A. Dhar, G. Mandal and S.R. Wadia, Int.J. Mod. Phys. \und{A8}
(1993) 325.
\bibitem{twelve} S.R. Das and A. Jevicki, Mod. Phys. Lett. \und{A5}
(1990) 1639.
\bibitem{thirteen} Handbook of Mathematical Functions, eds.
M. Abramowitz and I.A. Stegun, Chap. 9 (Dover, 9th printing, 1970).
\bibitem{fourteen} A. Dhar, G. Mandal and S.R. Wadia, `String beta
function equations from the $c=1$ matrix model', hep-th/9503172, to appear
in Nucl. Phys. B.
\bibitem{fifteen} C. Callan, D. Friedan, E. Martinec and M.
Perry, Nucl. Phys. \und{B262} (1985) 593.
\bibitem{sixteen} G. Mandal, A. Sengupta and S.R. Wadia, Mod. Phys.
Lett. \und{A6} (1991) 1685; E. Witten, Phys. Rev. \und{D44} (1991) 314.
\bibitem{seventeen} J. Polchinski, `On the nonperturbative consistency
of $d=2$ string theory', hep-th/9409168.
\end{thebibliography}
\end{document}